\documentstyle[12pt,aaspp4]{article}

\def\ni{\noindent}

\def\cm{{\rm\,cm}}

\def\gm{{\rm\,g}}

\def\au{{\rm AU}}
\def\AU{{\rm\, AU}}
\def\mum{\,\mu{\rm m}}
\def\K{{\rm\,K}}
\def\yr{{\rm\,yr}}

\def\th{\theta}
\def\tho{\theta_0}

\begin{document}

\lefthead{Chiang and Goldreich}
\righthead{PASSIVE T TAURI DISKS: INCLINATION}

\title{SPECTRAL ENERGY DISTRIBUTIONS OF PASSIVE T TAURI DISKS: INCLINATION}

\author{E.~I.~Chiang and P.~Goldreich}

\affil{California Institute of Technology\\
Pasadena, CA~91125, USA}

\authoremail{echiang@tapir.caltech.edu and pmg@nicholas.caltech.edu}

\begin{abstract}
We compute spectral energy distributions (SEDs) for passive T Tauri disks
viewed at arbitrary inclinations. Semi-analytic models of
disks in radiative and hydrostatic equilibrium are employed.
Over viewing angles for which the flared
disk does not occult the central star, the SED varies negligibly
with inclination. For such aspects, the SED shortward of $\sim$80$\mum$
is particularly insensitive to orientation,
since short wavelength disk emission is dominated by superheated surface
layers which are optically thin.
The SED of a nearly edge-on disk is that of a class I source.
The outer disk occults inner disk regions,
and emission shortward of $\sim$30$\mum$ is dramatically extinguished.
Spectral features from dust grains may appear in absorption.
However, millimeter wavelength fluxes decrease by at most a factor of 2
from face-on to edge-on orientations.

We present illustrative applications of our SED models.
The class I source 04108+2803B is considered a T Tauri
star hidden from view by an inclined circumstellar disk.
Fits to its observed SED yield model-dependent values for the disk mass
of $\sim$0.015 $M_{\odot}$ and a disk inclination of $\sim$65$^{\circ}$
relative to face-on. The class II source GM Aur
represents a T Tauri star unobscured by its circumstellar disk.
Fitted parameters include a disk mass of $\sim$0.050 $M_{\odot}$
and an inclination of $\sim$$60^{\circ}$.
\end{abstract}

\keywords{circumstellar matter --- radiative transfer --- stars: pre-main
sequence --- accretion, accretion disks --- stars: individual (GM Aur,
04108+2803B)}

\section{INTRODUCTION}
Excess infrared (IR) emission from T Tauri
stars is thought to originate from circumstellar disks
(\markcite{m68,sal87}Mendoza 1968;
Shu, Adams, \& Lizano 1987; and references therein). Passive disks are the
simplest
to consider. By definition, they lack intrinsic luminosity and
reradiate the energy they absorb from the central star.

Hydrostatic, radiative equilibrium models for passive T Tauri disks are derived
by \markcite{cg97}Chiang \& Goldreich (1997, hereafter CG).
The disk surface flares outward with
increasing radius and intercepts more stellar radiation, especially at large
distances from the central star, than a flat disk would
(\markcite{kh87}Kenyon \& Hartmann 1987).
An optically thin layer of superheated dust grains ensheaths the entire disk.
Dust grains
in this surface layer are directly exposed to central starlight and
reradiate to space about half the stellar energy they absorb. The
other half is emitted towards the midplane and regulates the temperature of the
cooler
disk interior. Vertical temperature gradients in externally illuminated
disk atmospheres are calculated in detail by \markcite{cetal91}Calvet et al.
(1991),
\markcite{mb91}Malbet \& Bertout (1991), and \markcite{daless98}D'Alessio et
al. (1998).

\markcite{cg97}CG compute spectral energy distributions (SEDs) of passive disks
viewed
face-on. The calculated SED is fairly constant
over the thermal IR, in accord with the observed flattish excesses of
T Tauri stars such as GM Aur.
Spectral features from dust grains in the superheated layer appear in emission
when the disk is viewed face-on, similar to solid-state emission lines
evinced by T Tauri stars and Herbig Ae/Be stars
\markcite{cw85,wetal96,ww98}(Cohen \& Witteborn 1985; Waelkens et al. 1996;
Waters \& Waelkens 1998).

What is the spectrum of a passive disk whose midplane is inclined at an
arbitrary angle, $\th$,
to the plane of the sky?
In \S\ref{contsed}, model assumptions and results pertaining to
the continuum SED are set forth.
In \S\ref{specdust}, we study how spectral signatures of some dust grain
resonances change from emission to absorption as the disk is viewed
increasingly
edge-on. Applications to observations are contained in \S\ref{appobs}.
There, we assess the possibility that the relatively low 12$\mum$ flux
of GM Aur might be caused
by a non-zero disk inclination, as opposed to a central AU-sized gap.
We close by considering whether
differences between class I and class II SEDs might reflect differences
in viewing angle rather than evolutionary status.\footnote{In the
classification
scheme of \markcite{lw84}Lada \& Wilking (1984; see also \markcite{l87}Lada
1987),
class I sources exhibit SEDs which rise from 2 to 10$\mum$, i.e.,
$-3 \lesssim n \lesssim 0$ where $\nu F_{\nu} \propto \nu^{n}$.
For class II sources, $0 \lesssim n \lesssim 2$.}

\section{INCLINATION DEPENDENCE OF CONTINUUM SED}
\label{contsed}

\subsection{Model Assumptions}
\label{assume}

We consider a passive disk in radiative and hydrostatic balance around a
T Tauri star. The model employed is identical to that
derived in \S2.1 and \S2.3 of \markcite{cg97}CG. Symbols and values of free
parameters
are listed here in Table 1, and henceforth are used without explanation. We
measure position by cylindrical radius, $a$, and vertical distance above the
disk midplane, $z$.\footnote{The disk radius in astronomical units is denoted
by
$a_{\au}$.}

\placetable{freeparm}
\begin{deluxetable}{cll}
\tablewidth{0pc}
\tablecaption{Free Parameters\label{freeparm}}
\tablehead{
\colhead{Symbol}      & \colhead{Meaning} &
\colhead{Value}}

\startdata
$M_*$ & Stellar Mass & $0.5 M_{\odot}$ \nl
$R_*$ & Stellar Radius & $2.5 R_{\odot}$ \nl
$T_*$ & Stellar Effective Temperature & $4000 \K$ \nl
$\Sigma$ & Disk Surface Density & $10^3 \, a_{\au}^{-3/2} \, \gm \cm^{-2}$ \nl
$a_i$ & Inner Disk Radius & $6 R_* = 0.07 \AU$ \nl
$a_o$ & Outer Disk Radius & $23000 R_* = 270 \AU$ \nl
$\kappa_V$ & Dust Opacity at Visual Wavelengths\tablenotemark{a} \phm{.} & $400
\cm^2 \gm^{-1}$ \nl
$\varepsilon_{\nu}$ & Grain Emissivity & $\left\{ \begin{array}{ll}
                                1 & \mbox{if $\lambda \leq 2\pi r$} \\
 (\frac{2\pi r}{\lambda})^{\beta} & \mbox{otherwise} \end{array} \right.$ \nl
$r$ & Grain Radius & $0.1 \mum$ \nl
$\beta$ & Grain Emissivity Index & 1 \nl
\tablenotetext{a}{Absorption by dust grains is assumed to dominate the
continuum
opacity from visible through millimeter wavelengths.}
\enddata
\end{deluxetable}

Derived disk properties relevant to our present investigation are summarized as
follows. At each radius we distinguish two regions: the superthermal surface
layer which is directly exposed to light from the central star, and the
cooler, diffusively heated interior which the surface encases. Variables
evaluated
in the former region are denoted by a subscript $s$, while those in the
latter region carry a subscript $i$. Pertinent results from \markcite{cg97}CG
include the dust
temperature at the surface

\begin{equation}
T_{ds} \approx \frac{550}{a_\au^{2/5}} \, \K \, ,
\label{tds}
\end{equation}

\noindent and the gas density in the interior

\begin{equation}
\rho_{gi} = {1\over\sqrt{2 \pi}} \frac{\Sigma}{h} \exp \left( -{z^2 \over 2
h^2}
\right) \, .
\end{equation}

\noindent
Both the gas scale height, $h$, and interior temperature, $T_{i}$, take on
different forms depending on the vertical optical depth of the interior. Inside
84$\AU$, the interior is opaque to its own reprocessed radiation, and

\begin{mathletters}
\begin{equation}
T_{i} \approx  \frac{150}{a_\au^{3/7}} \,\K \, ,
\end{equation}
\begin{equation}
h/a \approx  0.04 \,a_\au^{2/7} \, .
\end{equation}

\ni
Between 84 and 209 $\AU$, the interior is optically thin to its own
radiation but still thick to radiation from the surface; here

\begin{equation}
T_{i} \approx  21\,\K \, ,
\end{equation}
\begin{equation}
h/a \approx  0.15 \,\left(\frac{a_\au}{84}\right)^{1/2} \, .
\end{equation}

\noindent
Finally, in the outermost regions of the disk,
the encased material is transparent to radiation from the surface, and

\begin{equation}
T_{i} \approx  21 \,\left(\frac{209}{a_\au}\right)^{19/45} \,\K , \\
\end{equation}
\begin{equation}
h/a \approx  0.23 \,\left(\frac{a_\au}{209}\right)^{13/45} \,.
\end{equation}
\end{mathletters}

Boundary conditions are as follows.
The superheated layer is located at $|z|=H\approx 4h$. Its visual optical
depth normal to its flared surface is equal to the angle $\alpha$ at which
stellar rays penetrate the disk (refer to Figure 3 of \markcite{cg97}CG):

\begin{equation}
\alpha \approx
 \frac{0.4R_*}{a}+a\frac{d}{da}\left(\frac{H}{a}\right)\, .
\label{alpha}
\end{equation}

\ni  Radially, the disk extends from the silicate
condensation boundary at $a_i$, to an outer radius, $a_o$, at which $H \approx
a$.

The occulting angle,

\begin{equation}
\tho \equiv \arctan {a_o \over H(a_o) } = 45^{\circ} \, ,
\label{iocc}
\end{equation}

\ni is a natural angle with which to compare the viewing angle $\th$.
For $\th\lesssim \tho$, both the central star and most of the
disk surface are in direct view. For $\th \gtrsim \tho$, star
and inner disk are blocked from view by the flared outer ``wall''.

\subsection{Interior SED vs. $\Theta$}
\label{intsed}

\begin{figure}
\vspace{-1in}
\plotone{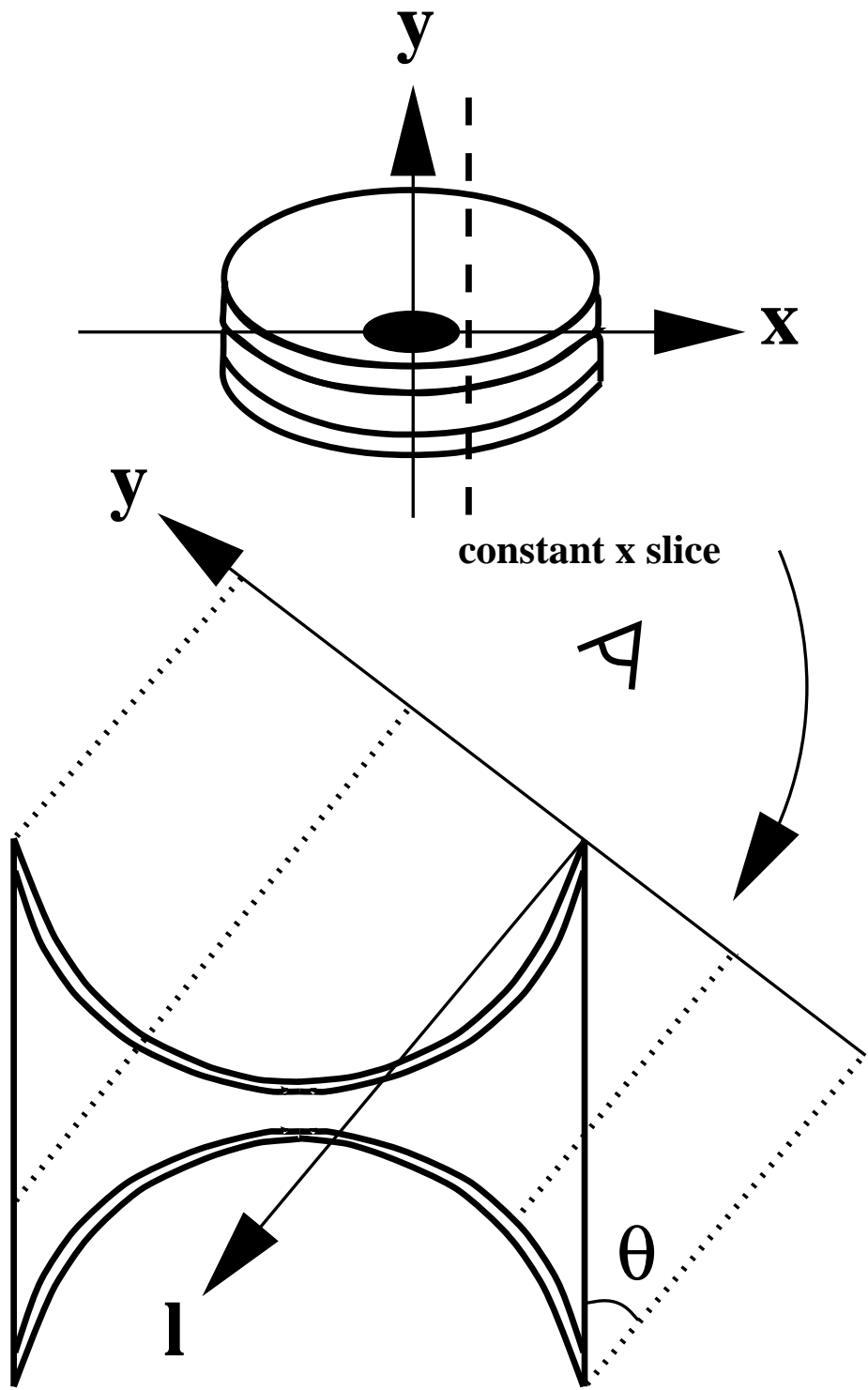}
\vspace{-1in}
\caption{Schematic of viewing geometry and coordinate system.
The disk interior is capped above and below by the superheated surface.
To calculate the SED, specific intensities along lines-of-sight
(dotted lines parallel to $l$) are summed over the projected
disk area in $x$ and $y$. Here $\th$ is slightly less than the
occulting angle $\tho$.\label{incschem}}
\end{figure}

\placefigure{incschem}

For clarity, we first consider only the disk interior.
Figure \ref{incschem} depicts the viewing geometry and coordinate system.
Define $x$ and $y$ to be orthogonal spatial axes in the plane of the sky,
centered on the star. The SED
is the integral of specific intensity over projected disk area, viz.,

\begin{equation}
L_{\nu} \equiv 4\pi d^2\nu F_{\nu} = 8\pi\nu \int_{0}^{a_o}\, dx\,
\int_{-y(x)}^{y(x)}\, dy\, I_{\nu}\, ,
\label{seddef}
\end{equation}

\ni
where

\begin{equation}
y(x) = \sqrt{a_o^2 - x^2} \cos \th + H(a_o) \sin \th \,
\label{yofx}
\end{equation}

\ni
traces the disk boundary on the sky, $d$ measures distance to the
source from Earth, and $I_{\nu} (x,y)$ is the specific
intensity;

\begin{equation}
I_{\nu} = I_{\nu, i} = \kappa_V \varepsilon_{\nu} \int_{0}^{\infty}\, dl\,
\rho_{gi}\, B_{\nu}(T_i)\, \exp{(-\tau _{\nu})} \, ,
\label{specint}
\end{equation}

\ni
with

\begin{equation}
\tau _{\nu} = \kappa_V \varepsilon_{\nu} \int_{0}^{l}\, d\tilde{l}\, \rho_{gi}
\, .
\label{taudef}
\end{equation}

\ni
Here $B_{\nu}$ is the Planck function, and $l$, $\tilde{l}$ both measure
line-of-sight distance from the observer.
Since temperature, density, and opacity are specified by our model, calculating
the
SED at arbitrary inclination is primarily an exercise in multidimensional
integration. We perform these integrals numerically using standard techniques
\markcite{press92}(Press et al. 1992).
Optical depths are evaluated
using a Romberg integrator with a mandated fractional accuracy of $10^{-3}$.
Specific intensities are computed on an adaptive
stepsize grid, where the source function ($B_{\nu}$) changes by no more
than 20\% between adjacent grid points. Integrals over $y$ are performed
using either a Romberg integrator or, in cases where the integrand is very
sharply peaked and the Romberg integrator
takes prohibitively many steps, a 60-point Gaussian quadrature routine.
The final integration over $x$ employs the trapezoidal rule
on a 60-point logarithmic grid. Answers are routinely checked for robustness
against changes in stepsize. We verify that the SED for
$\theta = 0^{\circ}$ computed using
the multi-dimensional integration code matches the SED for
face-on disks computed using the 1-dimensional integrator of CG.

\begin{figure}
\plotone{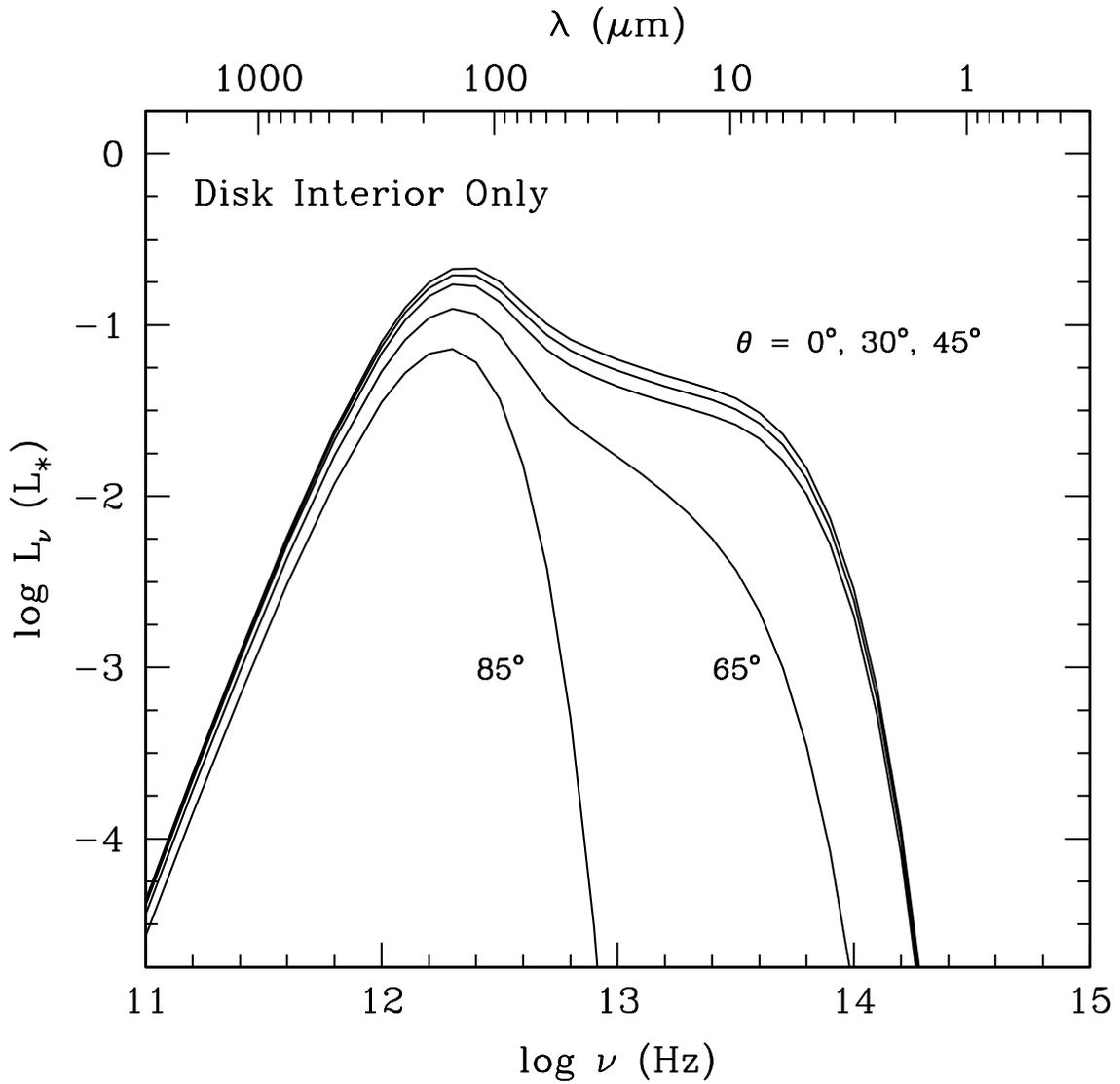}
\caption{SED of the disk interior only, as a function of viewing angle.
Short wavelength emission from the inner disk is extinguished by the
intervening outer disk for $\th > \tho$. $L_{\ast}$ denotes the luminosity of
the central star. \label{intersed}}
\end{figure}

Figure \ref{intersed} displays a family of interior SEDs for different $\th$.
First
consider SEDs for angles $\th \leq \tho \, (\th = 0^{\circ}, 30^{\circ},
45^{\circ})$. Radiation at wavelengths shorter than $\sim$100 $\mum$
comes mainly from optically thick regions; consequently, $L_{\nu}$ is
approximately proportional to the areal projection factor $\cos \th$.
At longer wavelengths, the radiation is emitted by increasingly transparent
material, so that $L_{\nu}$ tends to be independent of orientation.

\placefigure{intersed}

For angles $\th > \tho \, (\th = 65^{\circ}, 85^{\circ})$,
short wavelength emission from the inner disk is strongly extincted by
the intervening outer disk. However, millimeter (mm) wavelength fluxes only
drop by
a factor of 1.8 as the disk is viewed increasingly edge-on, since about half of
the outer disk remains visible in that limit.

\subsection{Superthermal + Interior SED vs. $\Theta$}
\label{supintsed}

Next we compute the total SED. The superheated layer is
treated as a plane-parallel atmosphere having
visual optical depth $\alpha$ normal to the surface.
Whenever a line-of-sight intersects the surface, we increment
the specific intensity from the interior by\footnote{This prescription fails
for lines-of-sight which graze the surface tangentially. However, their
fractional contribution to the integrated flux is of order
$\sqrt{h_s/R}$, where $h_s\approx (\partial \ln \rho_{gi} / \partial
z)^{-1}|_{z=H}$ is the geometrical thickness of the surface, and $R$ is the
radius
of curvature of $H$. Since this fraction is less than $10\%$, for computational
simplicity we let the intersected optical depth of the surface be $\alpha
\varepsilon_s
\times \min( |\hat{n} \cdot \hat{l}|^{-1} , 20)$.}

\begin{equation}
\Delta I_{\nu} = B_{\nu}(T_{ds}) \, \left[1 - \exp\left(-{\alpha \varepsilon_s
\over |\hat{n} \cdot \hat{l} |}\right) \right] \, \exp(-\tau_{\nu,l}) \, .
\end{equation}

\ni Here, $\hat{n}$ and $\hat{l}$ are unit vectors normal to the surface
and parallel to the line-of-sight, respectively, $\varepsilon_s$ is the
Planck-averaged dust emissivity at the surface, and $\tau_{\nu,l}$ is the
intervening optical depth between disk surface and observer.

\begin{figure}
\plotone{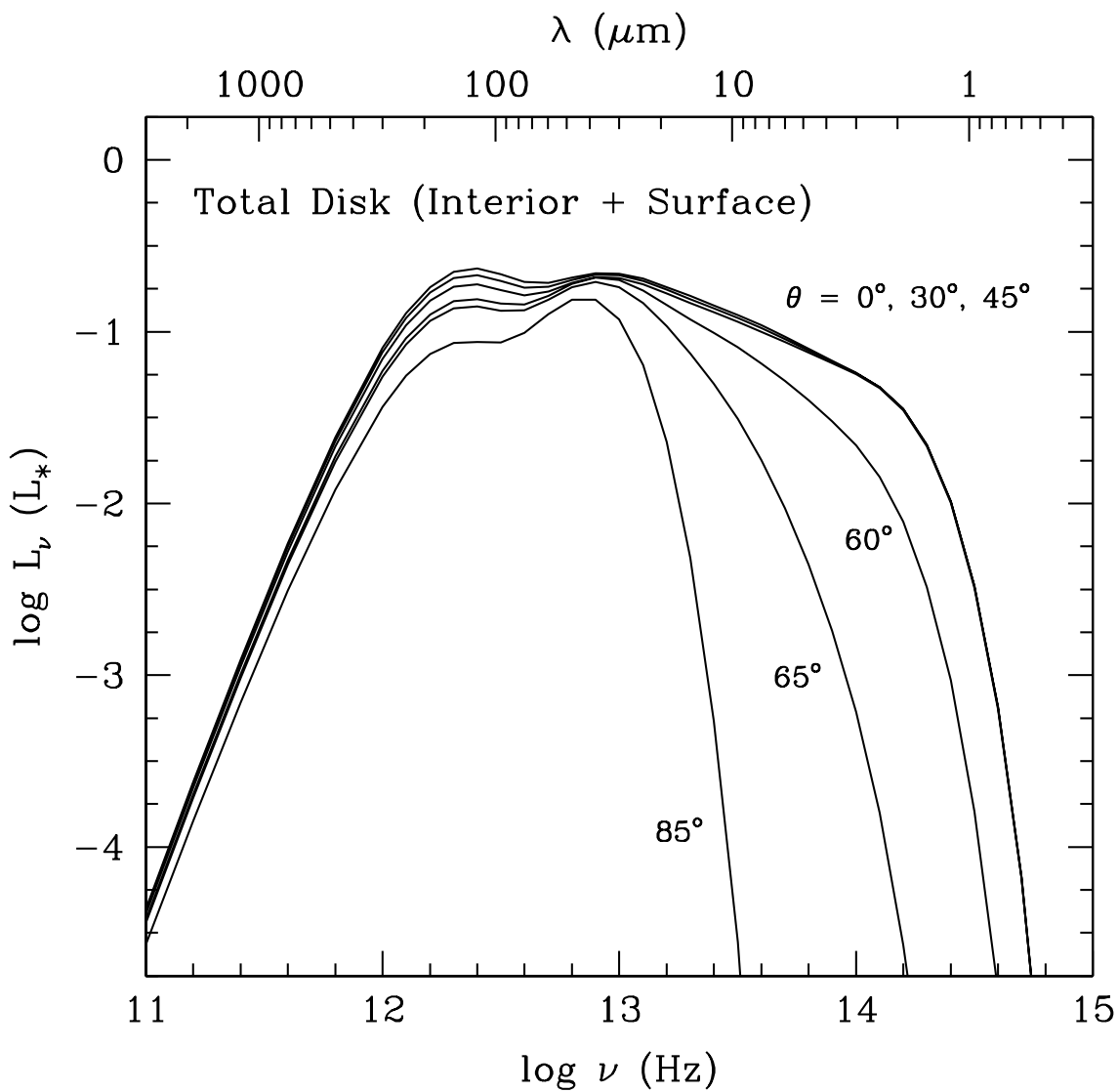}
\caption{SED of the entire disk (interior + surface),
as a function of viewing angle. For $\th < \tho$, the SED
is insensitive to viewing aspect since short wavelength radiation
originates from the optically thin surface, while long wavelength
radiation is emitted by the optically thin interior.
For $\th > \tho$, the intervening outer disk dramatically extinguishes
surface radiation at the shortest wavelengths. \label{totsed}}
\end{figure}

\placefigure{totsed}

Figure \ref{totsed} displays a family of total disk SEDs labeled by $\th$.
The behavior of the SED for $\lambda \gtrsim 80 \mum$
is similar to that described in \S \ref{intsed}, since the
interior dominates emission at those wavelengths.

Shortward of $\sim$$80 \mum$, emission from the superheated surface
makes a qualitative difference to the appearance of the SEDs.
One difference is that for angles $\th \leq \tho$, the flux
is remarkably insensitive to viewing geometry.\footnote{Slight reductions in
mid-infrared
flux still occur for $\th \leq \tho$, both because surface layers
intermediate in radius are seen through the outer disk's atmosphere and
because emission from the optically thick interior is proportional to
the projected area.
}
This is because shorter wavelength radiation emerges primarily from
the superheated surface layers, which are optically thin along most lines of
sight.

For angles $\th > \tho$, the intervening outer disk dramatically extinguishes
surface radiation at the shortest wavelengths. However, the total disk SED
between 30 and 80$\mum$ is relatively robust to changes in inclination;
it decreases by less than a factor of 3 between face-on and edge-on
orientations. Radiation at these wavelengths originates from surface
layers at large radii and is only
slightly obscured by the tenuous interior at still larger radii.

\begin{figure}
\plotone{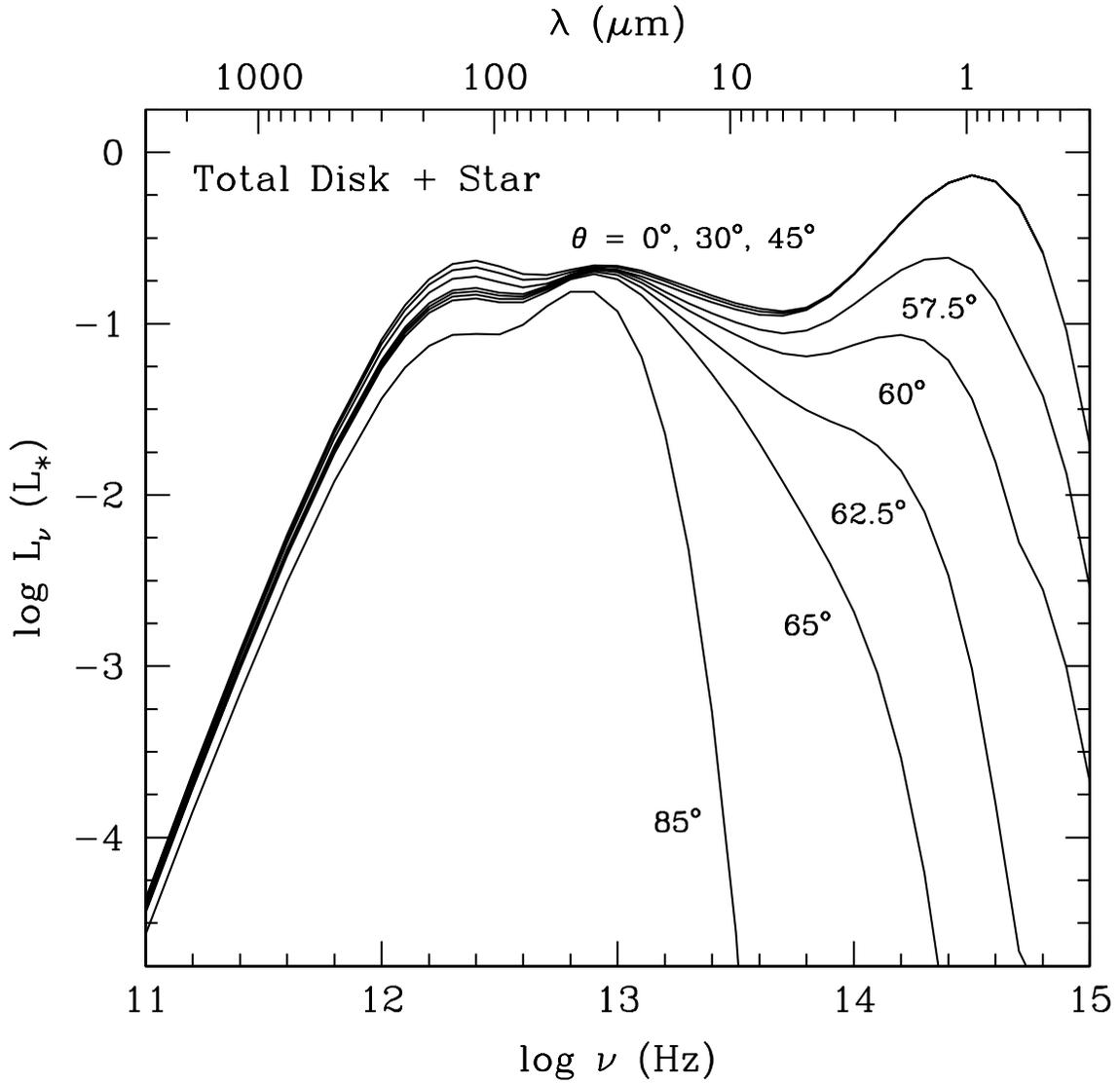}
\caption{SED of the entire disk + central star, as a function of
viewing angle. At $\th = 60^{\circ}$, the visual optical depth to the star
is $\tau_{\ast ,V} = 4.5$. At $\th = 65^{\circ}$, $\tau_{\ast ,V} = 22$.
At $\th = 85^{\circ}$, $\tau_{\ast ,V} = 7.6 \times 10^5$.
\label{starsed}}
\end{figure}

Finally, in Figure \ref{starsed}, we add the contribution from the central star
to the SED. The stellar flux fades rapidly with increasing $\th$
beyond $\sim$$55^{\circ}$.

\placefigure{starsed}

\section{Spectral Dust Features}
\label{specdust}

When viewed nearly face-on, passive disks exhibit emission lines associated
with dust grain resonances. These features arise from material along the
line-of-sight which is optically thin: line-to-continuum flux ratios
are of the same order as the percentage enhancement in line opacity.
Emission lines may originate from both the optically
thin superheated surface (as discussed in \markcite{cg97}CG) and the optically
thin
interior.

How do such lines vary as the disk is viewed increasingly edge-on?
Might they appear in absorption instead, as
\markcite{cw85}Cohen \& Witteborn (1985) find
for 7 of the 32 T Tauri stars they survey spectrophotometrically?
To address these questions, we add 6 Gaussian-shaped spectral
resonances to our grain emissivity law. In choosing resonant
wavelengths and strengths, we aim to illustrate the range of effects which
occur
with variable viewing angle. Roughly, our 6 lines can be associated with
crystalline H$_2$O ice at 60 and 100$\mum$
\markcite{ww98}(Waters \& Waelkens 1998), amorphous silicates
at 10 and 20$\mum$ \markcite{m90}(Mathis 1990), and
small carbon-rich grains or PAHs (polycyclic aromatic hydrocarbons) at
3.3 and 6.2$\mum$ \markcite{wetal96}(Waelkens et al. 1996).

\placefigure{specline}
\begin{figure}
\plotone{f5.port.epsi}
\caption{Solid-state line spectra at various disk inclinations.
Resonances at the shortest wavelengths where opacities are highest
appear in absorption over a limited range of viewing angles between $\tho$ and
$90^{\circ}$. Longer wavelength features persist in emission.
\label{specline}}
\end{figure}

Solid-state line spectra at various disk inclinations are displayed
in Figure \ref{specline}.
Emission at all wavelengths varies little with viewing geometry for angles
$\th\lesssim \tho$.
This is true particularly at the
shortest wavelengths where lines arise largely from the optically
thin surface layers. At wavelengths progressively longer than $\sim$30$\mum$,
the optically thin interior at large radius contributes increasingly
to line emission. The drop in the long wavelength continuum with increasing
inclination reflects the $\cos\th$ dependence of radiation emitted by the
optically thick interior.

Emission lines at $\sim$60 and $100 \mum$ persist at angles $\th \gtrsim \tho$,
although their fluxes are lower by a factor $\sim$2 for edge-on as
compared to face-on disks. For edge-on orientations, only about half of the
optically thin interior remains visible. Emission around $20 \mum$
is more robust to changes in inclination; most of this radiation
emerges from the superheated surface at large radii and passes through
the rarefied disk interior at yet larger radii with little attenuation.

At the shortest wavelengths, where occultation of the inner disk is
significant, lines indeed appear in absorption, though only over a
limited range of inclinations between $\tho$ and 90$^{\circ}$.
For a dust grain resonance to appear in absorption, disk inclinations must be
sufficiently high that line emission from the superheated surface is
extinguished. However, $\th$ cannot be so high that line and continuum optical
depths both exceed unity at large radius where the disk interior
is nearly isothermal. Only over an intermediate range of inclinations
does radiation on and off the resonant wavelength probe a variety
of interior disk temperatures. For the $3.3 \mum$ resonance to
appear in absorption, $60^{\circ}\lesssim i\lesssim 65^{\circ}$.
At $6.2 \mum$, opacities are lower, and inclinations needed for absorption
higher: $65^{\circ}\lesssim i \lesssim 70^{\circ}$.
In principle, near-infrared line spectra may provide a sensitive diagnostic of
disk inclination.

While the trends outlined above apply qualitatively to all passive disks,
quantitative conclusions are model-dependent. For example, in our standard
model, no absorption line at $10 \mum$ appears because line-of-sight
optical depths are too small to extinguish emission from the
superthermal surface. However, such an absorption line would appear in the SED
of a sufficiently massive, inclined disk.

\section{Applications to Observations}
\label{appobs}

\subsection{GM Aurigae and Central Holes}
\label{gmhole}
GM Aur is a particularly clean, well-studied T Tauri system to which
we can apply our SED models. The single central star has mass $0.72 M_{\odot}$,
luminosity $0.7 L_{\odot}$, and age $2 \times 10^6 \yr$
\markcite{betal90}(Beckwith et al. 1990).  Aperture synthesis maps
in $^{13}CO (2\rightarrow 1)$ evince a rotating
circumstellar gas disk having approximate projected dimensions 950 $\times$
$530$ $\au$, inclined at $\th \approx 30^{\circ}$
\markcite{ksb93}(Koerner, Sargent, \& Beckwith 1993). Hubble Space Telescope
(HST) images in scattered visible light reveal the disk surface to be flared,
and suggest $i$ is closer to $\sim$$60^{\circ}$
\markcite{k97,setal95}(Koerner 1997; Stapelfeldt et al. 1995).
These same observations detect no outflow or remnant envelope
surrounding this relatively evolved system.

\markcite{cg97}CG model the SED of GM Aur, but with a face-on disk. They
suggest that a
non-zero disk inclination
might cause line-of-sight radiation from the inner disk to be absorbed
by the outer disk, thereby explaining the relatively low $12\mum$ flux
without invoking a central hole. Here we investigate
this possibility by refitting the SED with an inclined disk.

\begin{figure}
\plotone{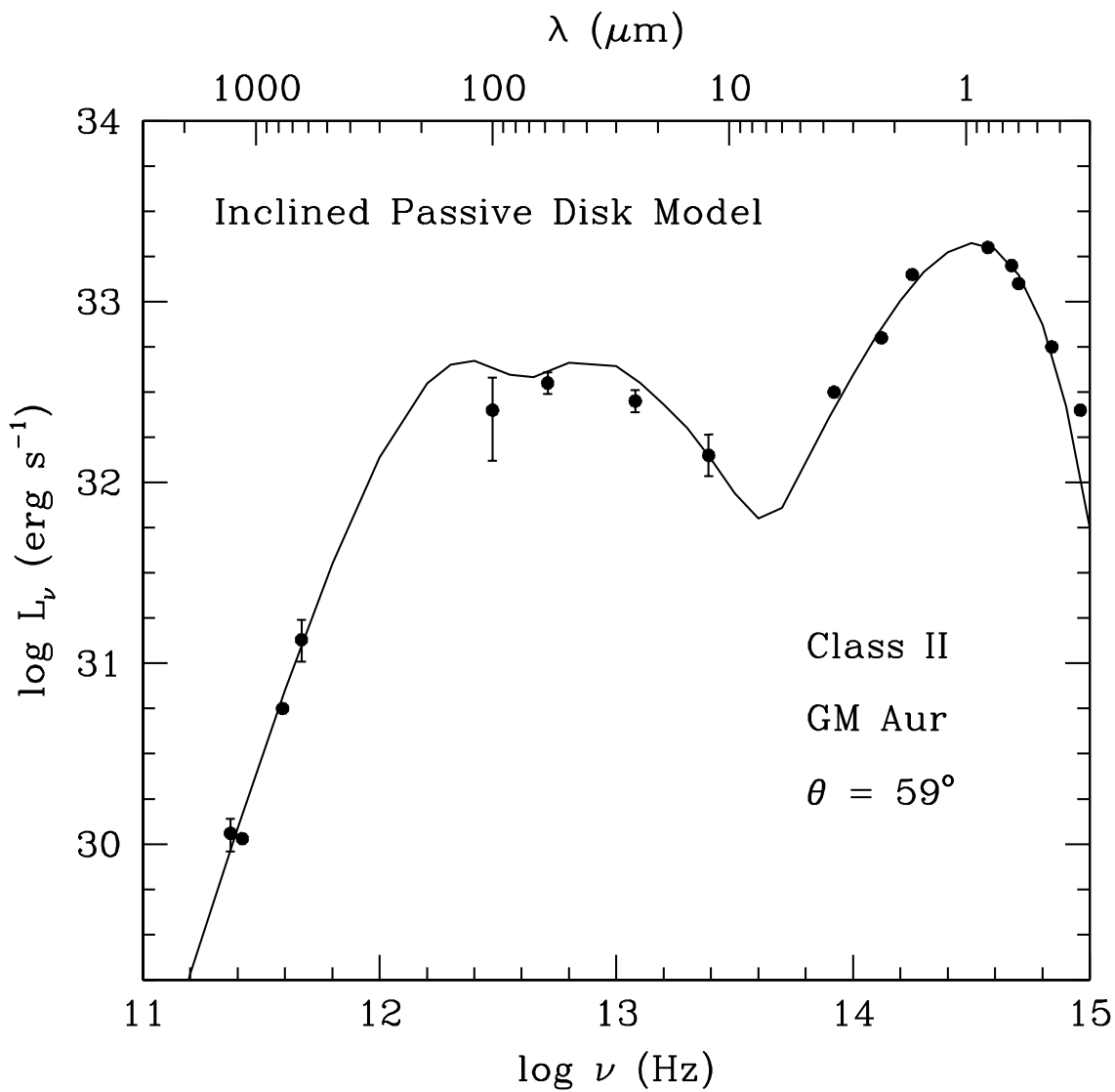}
\caption{Observed SED (filled circles) of GM Aur and accompanying
inclined passive disk model. Fit parameters are as follows:
$\th = 59^{\circ}$,
$\Sigma = 2 \times 10^3 \, a_\au^{-3/2} \, \gm \, \cm^{-2}$, $\beta = 1.4$, $r
= 0.3 \mum$,
$a_o = 311\AU$ (which implies $\tho = 51^{\circ}$), and $a_i = 4.8\AU$.
The formal requirement of a large AU-sized central
gap persists despite accounting for a non-zero viewing angle.
\label{gmsed}}
\end{figure}

\placefigure{gmsed}

Figure \ref{gmsed} displays our new fit with a disk inclined at $\th =
59^{\circ}$.
Despite the non-zero inclination, none of our fit parameters changes
significantly from those given by \markcite{cg97}CG for the face-on case.
Fluxes at IRAS
(Infrared Astronomical Satellite) wavelengths emerge mainly from the optically
thin surface layers, and those at mm wavelengths emerge mostly from the
optically thin
interior; neither is sensitive to viewing angles $\th$ near or less
than $\tho$. \footnote{~$\tho$ for this disk equals $51^{\circ}$.}

Our model reproduces the observed modest extinction to the central star
of $A_V = 0.5$ magnitudes, but at the cost of still requiring a large central
gap having a radius $\sim$$60$ times greater than the dust sublimation
radius: $a_i = 4.8 \pm 2.7 \au$, where the uncertainty reflects that of the
$12\mum$ IRAS point. There is no inclination for which the outer edge of our
model
disk attenuates the $12\mum$ flux without also extinguishing
central starlight by several magnitudes. Line-of-sight
column densities to surface regions inside $\sim$5 AU
(where most of the $12\mum$ emission originates) are only $\sim$$20\%$
greater than the column density to the star. This small
difference in the amount of obscuring material is completely insignificant;
the opacity at $12\mum$ is $13$ times lower than at visible
wavelengths. Thus, the low visual extinction to the central star
implies that we should have a clear view of disk regions
inside a few AU at near-infrared wavelengths.

In spite of these considerations, we remain skeptical of the
existence of such a large central gap devoid of dust.
A more palatable alternative might be that the inner disk's aspect
ratio does not increase monotonically with radius. Such undulations in
the height of the surface would shadow annular regions from the
central star and lower their temperatures. In addition, those ripples nearest
the observer would hide their warmer, starlit sides when viewed
at non-zero inclination. Accretional heating may be responsible
for such changes in surface geometry, as mentioned in \markcite{cg97}CG (see
their \S 3.5);
as estimated there, enhancements in disk thickness due to heating
of the midplane may become significant inside a few $\au$.
For an accretion disk which derives its luminosity solely from
local viscous dissipation, the aspect ratio indeed decreases with radius
whenever opacities increase steeply with temperature, as
demonstrated by \markcite{betal97}Bell et al. (1997).

\subsection{Class I Sources As Inclined Class II Sources}
\label{class12}
\begin{figure}
\plotone{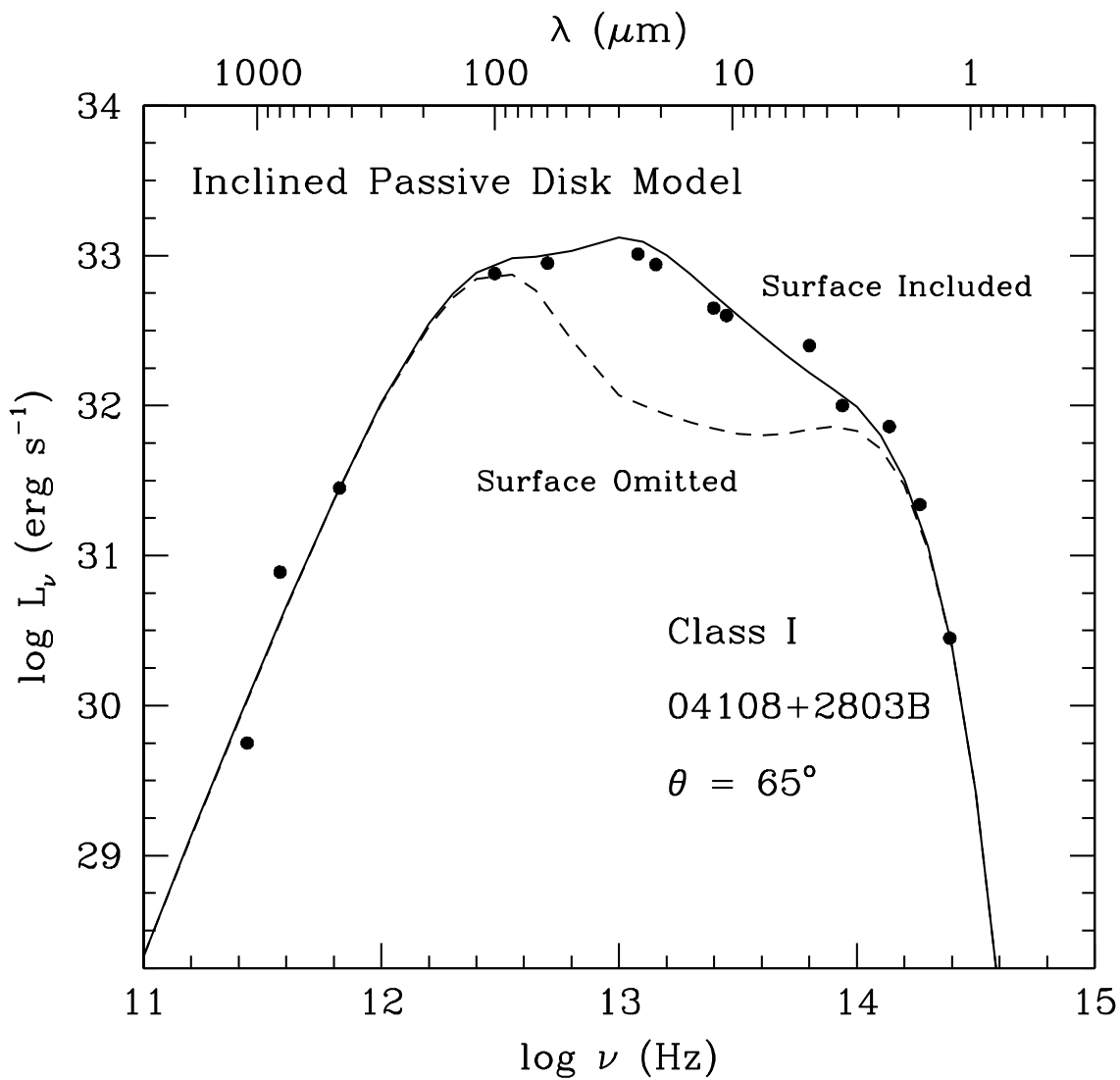}
\caption{Observed SED (filled circles) of class I source 04108+2803B
\protect{\markcite{kch93}}
(Kenyon, Calvet, \& Hartmann 1993)
and accompanying inclined passive disk model. Fit parameters for the disk
are as follows: $\th = 65^{\circ}$,
$\Sigma = 0.3 \times 10^3 \, a_\au^{-3/2} \, \gm \, \cm^{-2}$,
$\beta = 1.2$, $r = 0.1 \mum$,
$a_o = 270\AU$ (which implies $\tho = 46^{\circ}$), and $a_i = 0.07\AU$.
Stellar parameters are identical to those of our standard model T Tauri star.
\label{class1sed}}
\end{figure}

Might some class I spectra represent T Tauri stars obscured by
inclined disks? The possibility should be entertained for
sources such as 04108+2803B; Figure \ref{class1sed} demonstrates that a
passive disk inclined at $65^{\circ}$ provides a reasonable
fit (to within factors of 2) to this class I SED.
Fit parameters are similar to those of our standard model,
and are listed in the figure caption. For this particular model,
near-IR emission at J, H, and K is
interpreted as central starlight extincted by the disk's
outer edge; inclusion of starlight scattered off the disk surface
would imply a larger inclination. To highlight the contribution
from the superheated surface, we also plot the SED
with the surface emission removed.
Fluxes between 3 and 60$\mum$ arise primarily from
the superthermal surface, seen both through and over the lip
of the outer disk atmosphere (refer to Figure \ref{incschem}).

\placefigure{class1sed}

Clearly, a class I SED does not imply a unique distribution of
circumstellar material. High angular resolution images provide
additional clues. Many class I sources exhibit near-IR nebulosity
on scales ranging from 1500 to 3000 AU
\markcite{ketal93}(Kenyon et al. 1993; and
references therein); for these sources, an additional, non-disk
component of dust is needed. \markcite{kch93}Kenyon, Calvet, \& Hartmann (1993)
model class I SEDs using rotationally flattened,
infalling envelopes that are passively heated by central stars.
These envelopes extend out to larger radii than do our disks,
typically 3000 AU, but contain roughly the same amount of mass,
about 0.1$M_{\odot}$. Bipolar holes empty of material, presumably
evacuated by outflows, are invoked so that central starlight may
scatter off cavity walls towards Earth, thereby explaining
the observed near-IR fluxes. In imaging observations of three
class I sources using HST/NICMOS (Near Infrared Camera and
Multi-Object Spectrometer), \markcite{petal98}Padgett et al. (1998)
discover nearly edge-on, flared circumstellar disks having diameters
300--750 AU. Near-IR emission is observed to be scattered not only off upper
and lower disk surfaces, but also off ``dusty material within or on
the walls of outflow cavities.''

In general, a combination of an inclined,
passively heated disk, and a dusty bipolar outflow or partially
evacuated envelope may best describe class I sources.
Since 04108+2803B reveals no optical or near-IR emission beyond
$\sim$140 AU from
its central star \markcite{kch93}(Kenyon, Calvet, \& Hartmann 1993), it is
consistent with being a limiting example of a simple inclined disk.
Two other examples for which this limiting case scenario may also
apply include the embedded sources 04295+2251 and 04489+3042.
Both sources exhibit nearly flat excesses between $\sim$3 and
100 $\mum$ which may arise from flared circumstellar disks.
Neither source betrays extended near-IR emission on scales greater
than a few hundred AU, or high-velocity molecular gas from an outflow
\markcite{ketal93}(Kenyon et al. 1993).

Near-IR polarimetry also addresses the possible presence
of envelopes. \markcite{wkg97}Whitney, Kenyon, \& G\'{o}mez (1997)
determine that linear polarizations greater than $\sim$20\%
at J, H, and K characterize the largest reflection nebulae associated
with class I sources. Such large polarizations are
interpreted as arising from the scattering of central starlight off cavity
walls in envelopes. In contrast, 04108+2803B evinces fractional
polarizations of 5.1\% and 1.6\% at H and K, respectively
\markcite{wkg97}(Whitney, Kenyon, \& G\'{o}mez 1997)---levels
more comparable to those in class II sources than in truly younger, more
embedded protostars, and which may arise from starlight scattered
off the disk surface. Resolved polarimetry maps of 04108+2803B
can test our hypothesis.

Degeneracies inherent in models deduced from a SED may also be broken by
kinematic studies. Interferometric radial velocity maps in molecular
lines may distinguish between an infalling envelope
(e.g.,\markcite{hom93,ohhm97} Hayashi, Ohashi, \& Miyama 1993;
Ohashi et al. 1997), and a rotating disk (e.g.,\markcite{ks95}
Koerner \& Sargent 1995).

Finally, we note that the inverse problem to 04108+2803B is presented
by the star HK Tau B: a system observed to possess an edge-on
($i \approx 85^{\circ}$) circumstellar disk and no observable envelope,
but whose infrared SED is not well
measured because of confusion from an infrared-bright companion
\markcite{setal98,kor98}(Stapelfeldt et al. 1998; Koresko 1998).
We await SIRTF (Space Infrared Telescope Facility) which can
provide both the SED between 3 and 180 microns and also
images of the superheated disk surface in thermal emission.

\acknowledgments
Financial support for this research was provided by NSF grant 94-14232,
NASA grant NAG5-7008, and an NSF Graduate Fellowship. We thank an
anonymous referee for a careful reading of our paper and for providing
suggestions which helped to improve its presentation.

\end{document}